\documentclass[]{spie}  

\usepackage[]{graphicx}
\def\titlefont{\LARGE\bf} 
\title{Quantum key distribution with unconditional security for all optical fiber network} 

\author{
Osamu Hirota\supit{a,b}, 
Kentaro Kato\supit{b}, 
Masaki Shoma\supit{c},
and 
Tsuyoshi Sasaki Usuda\supit{d},
\skiplinehalf
\supit{a}
Tamagawa University, 
Tamagawa-gakuen 6-1-1, Machida, Tokyo 194-8610, JAPAN\\
\supit{b}
21st century COE program, Chuo University, Tokyo, JAPAN\\
\supit{c}
Matsushita Research Institute  Tokyo, Inc., JAPAN\\
\supit{d}
Aichi Prefectural University, Aichi, JAPAN.
}

\authorinfo{
Further author information: 
(Send correspondence to Osamu Hirota)\\
Osamu Hirota: address; 6-1-1, tamagawa-gakuen, machida-city, Tokyo, Japan.
E-mail, hirota@lab.tamagawa.ac.jp; 
Telephone, +81-42-739-8674\\  
Kentaro Kato: 
E-mail, kkatop@ieee.org\\
Masaki Sohma: 
E-mail, sohma@mrit.mei.co.jp\\
Tsuyoshi Sasaki Usuda: 
E-mail, usuda@ist.aichi-pu.ac.jp
}

\begin{document} 
  \maketitle 

\begin{abstract}
Previously the present  protocol was referred as Yuen-Kim second version in our papers. 
In this paper, it is called Yuen protocol (Y-00) and we present an efficient implementation method of 
physical layer of Y-00 which can support a secure 
communication and a quantum key distribution (more generally key expansion) 
by IMDD(intensity modulation/direct detection) 
or FSK(frequency shift keying)optical fiber communication network.
Although the general proof of the security is not yet given, 
a brief sketch of security analysis is shown, which involve an entanglement attack.
\end{abstract}



\section{INTRODUCTION}
A quantum key distribution for legitimate two users
(Alice and Bob) is one of the most
interesting subjects
in quantum information science, which was pioneered by C.Bennett and
G.Brassard in 1984[1]. In the original paper of Bennett, 
single photon communication was employed for implementation
of quantum key distribution for simple explanation. 
However, despite that it is not essential 
in the great idea of Bennett, many researchers employed
single photon communication scheme to realize BB84, B92[2]. 
There exist limitations of communications by single photon state 
or other quantum state(squeezed state and et al) with strong 
quantumness in the practical sense[3]. So 
it was discussed whether one can realize a secure key distribution 
guaranteed by quantum nature based on conventional light wave 
communication or not. One of the answers is that one  employs  
coherent state as transmitter state, which has a semiclassical nature and 
a robustness for energy loss. Recently, as an example, an implementation of 
BB-84 by using continuous variable coherent state has been proposed[4],  
which may communicate more long distance and higher efficiency than 
single photon scheme.

On the other hand, in the history of information theory, 
there were already several works on  protocol with unconditionally secure 
communication.  Let us introduce a short history on this subject. 
The most important problem is what is unconditional security.
The condition of  perfect secrecy $I(X;C)=0$ was given  by Shannon. 
It  means that the plaintext $X$ and the cipher text $C$ as a function of 
$X$ and a secret key $K$ should be statistically independent. 
He assumed in his model that Eve has access to precisely the 
same information as the legitimate user. 
As a result,  a condition: $H(X)\le H(K)$ is required. 
It looks like impractical, because it means that perfect secrecy is 
achieved only when the secret key is at least as long as the plaintext message. 
However, such a pessimism was solved by introducing the modified 
Shannon's model such that Eve cannot receive precisely the same information as Bob.  There may be certain condition in physical world to realize the perfect secrecy. If we have no conditions for it, then it is called unconditionally secure scheme.
In general, we can define a channel model of cipher communication or 
key distribution by the conventional information theory. 
That is, we have two channels. One is 
the channel from Alice to Bob, and the other is the channel from Alice to Eve. 
Let $X,Y,Z$ be random variables of Alice, Bob, and Eve, respectively.
The channel behavior is completely specified by the conditional probability 
$P(Y, Z | X)$.
For broadcast channels described by $P(Y, Z | X)$, Wyner[5] clarified a fact 
that one can realize a scheme with unconditional security in which Eve is assumed 
to receive signals from Alice over a channel that is noisier than 
the legitimate users. Subsequently Csisz$\acute{\rm}$ar-K$\ddot{\rm o}$rner[6] 
generalized Wyner's result defining the secrecy capacity $C_c(P(Y, Z | X))$ 
which means that the maximum bits that Alice can send to Bob in secrecy. 
These schemes allow us  realization  of one way communication with confidential 
message $X$ without initial key, and also distribution of key $K$. 
However, the assumption that Eve's channel is worse than the legitimate 
channel is also unrealistic. 
When the discussion is devoted to only the problem of  key distribution, 
Maurer[7] pointed out that this assumption is not needed if the legitimate 
users can communicate over insecure but authenticated public channel like BB-84.
Unfortunately the efficiency of communication is not so good, 
because it requires small SNR. In addition, in such information 
theoretical results, the problem of finding actual encodable and 
decodable codes that perform in a particular situation was remained.
Yuen and Kim[8] showed that, using  a protocol like B-92, a practical 
system with unconditional security  can be realized  
if there exist some statistically independent noise or quantum shot noise 
in receivers of Eve and Bob, even the system is conventional fiber communication.  
It is called YK protocol.  A simple experimental demonstration of 
YK protocol based on classical noise was reported[9]. 
Thus the unconditionally secure key distribution does not require 
a realization of single photon communications and also even quantum phenomena.
{\it These results verified that the key distribution with unconditional security 
is not proper matter of quantum theory}. So one can realize unconditionally  
secure scheme by classical communication systems if there exist unavoidable 
noise such as "thermal noise in free space" and so on. In addition, $H(X)\le H(K)$ 
is not essential.  
However, these  schemes are still inefficient, because it requires small SNR also.   
One should investigate protocol with more high rate under high security. 
The best way may be that one seeks noisy channel governed by unavoidable 
quantum noise  based on quantum mechanics. As a result, an interesting question arises
" Is it possible to create a quantum system with current technology that 
could provide a communication in which always Bob's error probability is 
superior to that of Eve?"
In 2000, Yuen gave a protocol(Y-00) to realize it as a positive answer[10], 
in which he clarified a new scheme  so called $M$-ary level cipher.
This provides a theoretical basis for communication with confidential message 
for more general channel models including Wyner's broadcast channel, 
but it requires initial shared key between Alice and Bob.
Furthermore, his scheme supports also secure key expansion,  
because the message can be replaced by random number.
In proceedings paper of QCM and C 2002, an implementation method was shown by using 
conventional optical communication system[11]. 
These results urged the problem of Wyner, Csisz$\acute {\rm a}$r, 
Maurer to return from classical to quantum. It may be expected that 
it will open a new trend of 
quantum cryptography as a replacement of single photon scheme.

Thus, the theory of quantum key expansion has been developed along two directions. 
One is to follow BB-84  and to give rigorous proof of security of BB-84 and B-92. 
The other is to follow a basic idea of Wyner and others for broadcast channels. Y-00  
is close to  the latter, but it is applicable to more general channels. 

In this paper, we will present an implementation  method of  physical layer for 
so called Y-00 protocol which supports a secure communication and 
a quantum key expansion  applicable to IMDD(intensity 
modulation and direct detection) or FSK(frequency shift keying) fiber 
communication network with fiber amplifiers 
as optical repeater, and propose an error correction method to improve 
the performance of the total system. 

\section{Yuen protocol:Y-00}
\subsection{Basis of protocol}
It is well known that the no cloning theorem is not necessary condition 
for unconditionally secure key expansion in information theoretical 
cryptography, and it is only specific example of the realization. 
The general information theoretical condition for unconditional security 
was discussed in the field of information theory[5,6]. Following this line, a physical cryptography by conventional optical communication was proposed 
by Yuen and Kim[8] which is called YK protocol.
Let us here briefly introduce a development from YK protocol to Y-00 protocol.  
A fundamental idea of YK protocol 
comes from the next remark.\\
{\bf Remark}: {\it If there are statistically independent noises between Eve
and Bob, there exists a secure key expansion protocol based on communication.}\\
This is an idea along the concept of {\it advantage distillation} of Maurer[7], 
and it is  regarded as a method 
to realize 
a scheme such that a channel of legitimate users is reliable than 
that of Alice and Eve, even the signal to noise ratio of Bob is less 
than that of Eve. 
Yuen and Kim emphasized that the essential point of security of the key distribution
comes from detectability of signals.
This is quite different from the principle of BB-84, et al which are
followed by the no cloning theorem.
That is, BB-84 and related protocol employ a principle of unavoidable disturbance 
to quantum states to guarantee the security,
but YK protocol employs a principle of communication theory to do so, 
and it can be realized as a modification of B-92 
which has reconciliation protocol through a public channel.
Although this can apply to long distance fiber system with optical repeaters, 
it cannot provide
unconditional security when we extend communication distance and in some 
practical situations. 
Unfortunately, all of protocols proposed so far including the above scheme is  
inefficient for applying to commercial use.
To cope with this defect, it was proposed [10] to use a fundamental theorem  
in quantum detection theory as follows[12]: 
\\
{\bf Theorem 1}: \\
{\it Signals with non commuting or the same density operators cannot be 
distinguished without error.}\\
This means that if we assign non commuting density operators for bit signals 1 and 0,
then one cannot distinguish without error.
When the error is 1/2 based on quantum noise, there is no way to distinguish them. 
So the problem is to create such a situation on
the channel between Alice and Eve, but the channel of Alice and Bob is 
kept as a normal communication channel. To realize it, Yuen employed a combination of  a  shared short  key  scheme for the legitimate users and a stream cipher  instead 
of reconciliation protocol. This realizes an advantage distillation.
A crucial new ingredient in his protocol is the explicit use of a shared short 
key for cryptographic objective of secure communication and key expansion.
We call it Yuen protocol(Y-00).  

Let us here give a short survey of original protocol and physical layer of Y-00 as direct encription.
{\em
\begin{itemize}
\item[\rm(a)] The sender(Alice) uses an explicit key(seed key:$K$,
expanded into a long random running key:$K^*$
by use of pseudo random generator as a stream cipher) to modulate the parameters of a multimode
coherent state.
\item[\rm(b)] State
$|\Psi_0\rangle=|\alpha/\sqrt{2}\rangle_1\otimes|\alpha/\sqrt{2} \rangle_2$
is prepared. Bit encoding can be represented as follows:
\begin{equation}
|\Psi_b\rangle=\exp\{-iJ_z\phi_b\}|\Psi_0\rangle
=|e^{-i\phi_b/2}\alpha/\sqrt{2}\rangle_1
\otimes|e^{i\phi_b/2}\alpha/\sqrt{2} \rangle_2
\end{equation}
where $J_z =({a^\dagger}_1{a_1} -{a^\dagger}_2{a_2})/2$.
\item[\rm(c)] Alice uses the running key $K^*$ to specify a basis from a set
of M uniformly
distributed two-mode coherent state.
\item[\rm(d)] The message $X$ is encoded as $Y_{K^*}(X)$. This mapping of
the stream of bits is the key to be
shared by Alice and Bob. Because of his knowledge $K^*$, Bob can demodulate from
$Y_{K^*}(X)$ to $X$.
\end{itemize}
}
In order to realize secure communication or to expand the key stream under secure way, 
we have to guarantee that Eve cannot obtain any information from the legitimate channel.
They applied that the ciphering angle $\phi_\nu$ could  have $k$ in general 
as discrete or continuous variable determined by running key generated from 
pseudo random generator. A ciphered two mode state may be
\begin{equation}
|\Psi_{bk}\rangle=\exp\{-iJ_z(\phi_b+\phi_k)\}|\Psi_0\rangle
\end{equation}
The corresponding density operator for all possible choices of $k$ is
$\rho_b$, where $b=1$ or $0$, where $\rho_b$ are density operator for Eve, 
and mixed state consisting of 
two-mode coherent states.
The problem for the security is to find the minimum error probability 
that Eve can achieve in bit determination. To find the optimum detection process for discrimination 
between $\rho_1$ and $\rho_0$ is a problem of quantum detection theory. 
The solution is given by [12]
\begin{equation}
P_e= \min_{\Pi}(p_1Tr\rho_1\Pi_0 + p_0Tr\rho_0\Pi_1)
\end{equation}
Here, $\{\rho_i\}$ are pure states for Bob, and are mixed states for Eve who has no initial key.
As an example of encoding to create $P_e(E) \rightarrow 1/2$ which is the error probability of Eve, 
Yuen et al., used a scheme such that closest values of a given $k$ can be 
associated with distinct bits from 
the bit at position  $k$, and two closest neighboring states represent distinct bits which means a set
of basis state. In this scheme, they assumed that one chooses 
a set of basis state(keying state for 1 and 0) for bits
without overlap. As a result, the error probability for density 
operators $\rho_1$
and $\rho_0$ becomes 1/2,
when the number $M$ of a set of basis state increases.
Asymptotic property of the error probability depends on the amplitude(or energy) of coherent state. 
So one requires large number of $M$ or modes when one wishes to extend the communication distance. 
Thus in this scheme Bob has always a better channel than Eve. This means the advantage distillation.

\subsection{Security analysis}
We clarify an intuitive proof of the security for Y-00 given by Yuen[23].
The security analysis of Y-00 is considered separately for 
secure direct encryption and key expansion.
For direct encryption, the seed key:$K$ is fixed and the running key:$K^*$ to modulate the optical modulator  is 
constructed by blocks of bits of pseudo random number generated(PRG) by 
pseudo random number generator with the seed key. 
We assume that the PRG has sufficient properties for uniformness and so on. In this scheme, Eve cannot get information bits, because of from Eq(3). But she can try to get directly key information through the partial information of running key based on her observation of quantum states as basis in modulation. 
This corresponds to known plain-text attack 
in the conventional theory, because she attacks based on running key information which get from her observation for quantum states. 
The problem is how much accuracy for the observed data is possible for getting "running key".  Since Eve has no a priori information on properties of random sequence as running key, she cannot apply several detection scheme known as correlation detections. So she has to employ optimum single shot detection scheme to get the most accurate data of each bit sequences. Let us consider a detection scheme based on individual quantum measurement and classical collective(or correlation) attack.
We can evaluate the security by the  quantum detection theory for quantum state signals which was formulated by Holeve and Yuen[12].
In our case, the detection limit is formulated by quantum minimax formula  of Hirota-Ikehara[24].\\
{\bf [Quantum minimax game between Alice and Eve]}\\
\begin{equation}
{P^*}_e = \min_{\Pi}\max_{p_i} (1 - \sum p_iTr \rho_i\Pi_i)
\end{equation}
where $\rho_i$ are quantum states which represent the basis state in quantum M-ary cipher,  $\{\Pi\}$ is positive operator valued measure(POVM) which correspond to receiver of Eve, $p_i$ is randomization probability of Alice. Finally the accuracy of Eve's data is given by the minimax value of this game.
By this ultimate detection limit, Eve has to estimate 
the running key based on her data which involves error.
If Eve can get the perfect information on running key,
then the security level of this scheme is the same as the conventional stream cipher. 
Let $n$ be length of the seed key. The security level of the ideal stream cipher 
is represented by $2^n-1$. 
In general, the security level of conventional stream cipher is weaker  than it. 
However, by quantum  M-ary cipher, the data of Eve involve inherent 
error based on quantum noise.
Here, if the number $M$ is enough large, the  equation (4) becomes ${P^*}_e  \cong 1$.
That is, the sequences of running key are truly randomized. Hence nobody can get the information on 
the correlation and others among the running key sequences from the observed data.
So the legitimate users have {\it almost secret channel} which is guaranteed by the ultimate quantum detection limit.

Thus, the security of classical stream cipher is enhanced 
by quantum  $M-$ary cipher scheme. As a result, Eve cannot apply the conventional correlation attack and others for the structure of the generator. The possibility for  Eve's strategy is only pure guess for all kind of key $2^n-1$ which corresponds to Brute force attack, even the security of pseudo random number generator 
is weak.  However, if Eve takes the strategy of Brute force attack, then she needs complete cipher text. 
A trial to get cipher text is equivalent to the cipher text attack in this case. The problem is  
how to do such a cipher-text attack.
First Eve has to clone the following sequence of quantum states
\begin{equation}
|\Psi_1 \rangle = (|\alpha_{i} \rangle_1|\alpha_{j} \rangle_2 |\alpha_{k} \rangle_3\dots  
\end{equation}
then this sequence is copied as follows:
\begin{eqnarray}
\begin{array}{lcl}
|\Psi_1 \rangle &=& (|\alpha_{i} \rangle_1|\alpha_{j} \rangle_2 |\alpha_{k} \rangle_3\dots  \\
|\Psi_2 \rangle &=& (|\alpha_{i} \rangle_1|\alpha_{j} \rangle_2 |\alpha_{k} \rangle_3\dots  \\
|\Psi_3 \rangle &=& (|\alpha_{i} \rangle_1|\alpha_{j} \rangle_2 |\alpha_{k} \rangle_3\dots  \\
\vdots
\end{array}
\end{eqnarray}
and all sequences must be storaged in quantum memory during the trial to check all kind of text.
However, each quantum bits are non orthogonal $\langle \alpha_{i}|\alpha_{j}\rangle \cong 1$, 
and also $\langle\Psi_i|\Psi_j \rangle \ne 0$.
So the clone has error due to quantum no cloning theorem when $M$ is enough large.  As a result, Eve cannot get exact cipher-text. 
{\it If the error is enough large, it is difficult to apply
the cipher-text attack, and also Brute force attack, because 
they would need a super exponential search.}
Thus, this scheme has almost perfect secrecy in the individual quantum measurement.  
For the case of the quantum collective measurement which will provide the same result, we will show in elsewhere. 

In the case of key expansion, the physical layer is the same as the case of direct encryption.
The information data is replaced into the true random number, 
and they are divided to the blocks with length $n$.
After communication of first blocks with length $n$, the legitimate users keep the block, 
and the seed key is refreshed by the bit block.
Then the next random number with the length $n$ is sent by the same physical process 
with refreshed seed key. 
They keep them and repeat the process. The bit strings(key)  accumulated by 
the above communication are used as one time pad cipher.
It is difficult to get the information for the block bit, because of 
the same reason for the case of the direct encryption.
Since the seed key for M-ary cipher is always refreshed, this is stronger than the case of direct encription.
Even if Eve can get some information in this case, the legitimate user can eliminate it by the privacy amplification. 
So finally  the unconditional security for quantum individual, collective measurements, and classical collective attack will be shown.

\section{Entanglement attack}
Analysis of security is an essential requirement for the discussion on cryptography.
Various eavesdropping strategies have appeared in security analysis for BB-84, B-92[13-15].  
A recent review is given by Gisin et al[16]. 
However, these are not always effective for other scheme, and depend on protocols. 
Here we discuss on the security for the present model.
There is no way for getting bit information  in the scheme, 
because the error of Eve is always 1/2 for bit sequence.  
However, in quantum system, Eve can  attack by methods 
as not only passive processing,  but also active processing.
For free space communication, 
one does not need to consider opaque attack and others, because the strategy of Eve is 
only to receive a piece of light wave from Alice. On the other hand, 
for fiber system, one has to consider various eavesdropping strategies 
in the same way as  BB-84.  One of them is entanglement attack. 
Although there are several variation of entanglement attack, 
here it is sufficient to consider the most simple one.
That is, since the legitimate users do not use public channel in Y-00, 
we do not consider general attacks as conventional collective and joint attack. 
The entanglement attack in this case is that Eve entangles a separate probe 
with coherent states from Alice and detect the probe.
After entanglement operation,  Eve has to
send the mode A which comes from Alice and keep the mode B as probe.
When Bob measures them, some effects come from the action. Even if the effects act 
completely, Eve cannot get information of bit, because the each state has 
possibility of 0 and 1(see section 4). It helps Eve only for information of quantum states 
which was sent by Alice.  If Eve can know all quantum state used by 
the legitimate users by such an entanglement attack, it reduces to classical stream cipher system. 

Here we show that Eve also cannot get information of quantum state used by the legitimate users.
To give a simple theory of entanglement for coherent state is worth while, 
because all information is transmitted by coherent state.
Let ${\cal H}_A$, and $ {\cal H}_T$ be the Hilbert spaces of Eve and 
of the total signal plus probe system, respectively.
If $|\alpha_i \rangle, | \phi \rangle, U$ denote the signal, 
probe's initial states and the unitary interaction, respectively, 
then the state of the signal received by Bob is given by 
the density operator obtained by tracing out Eve's probe:
\begin{equation}
\rho_{B} = Tr_{E}(U |\alpha_i, \phi \rangle \langle \alpha_i, \phi |U^{\dagger})
\end{equation}
The unitary interaction operated by Eve should make the largest entanglement 
between the signals and probes. Hence it is reasonable to discuss 
the general properties of entanglement for coherent states.
One of authors and van Enk gave a general analysis for entangled state of 
non orthogonal state like coherent state[17, 18]. Here we apply them to 
an evaluation of the security. 
The following quasi Bell states or unitary equivalence of them have good 
property for the entanglement.
\begin{eqnarray}
\left\{
\begin{array}{lcl}
|\Psi_1 \rangle_{AB}  &=& 
h_{1} (|\alpha \rangle_A|-\alpha \rangle_B 
+|-\alpha \rangle_A|\alpha \rangle_B ) \\
|\Psi_2 \rangle_{AB}  &=& 
h_{2} (|\alpha \rangle_A|-\alpha \rangle_B
-|-\alpha \rangle_A|\alpha \rangle_B )\\
|\Psi_3 \rangle_{AB}  &=& 
h_{3} (|\alpha \rangle_A|\alpha \rangle_B 
+ |-\alpha \rangle_A|-\alpha \rangle_B )\\
|\Psi_4 \rangle_{AB}  &=& 
h_{4} (|\alpha \rangle_A|\alpha \rangle_B 
-|-\alpha \rangle_A|-\alpha \rangle_B)
\end{array}
\right. 
\end{eqnarray}
where $h_i$ is normalized constant,  $\alpha$ is coherent amplitude of light field. 
As remarkable property, $|\Psi_2 \rangle_{AB}$ 
and $|\Psi_4 \rangle_{AB}$  have 
complete entanglement which is independent of $\alpha$. 
The entangled states of other type have bad entanglement.
If the amplitudes of mode A and mode B  in the above form are different, 
the eigenvalues of the reduced density operator become
\begin{equation}
\lambda_{1} = 
\frac{(1+\kappa_A)(1-\kappa_B)}{2(1-\kappa_A \kappa_B)}, 
\quad 
\lambda_{2} = 
\frac{(1-\kappa_A)(1+\kappa_B)}{2(1-\kappa_A \kappa_B)}
\end{equation}
where $\kappa_A$ and $\kappa_B$ are inner products of basis states in mode A and B, 
respectively. As a result, again we can see that degree of entanglement is 
maximum when $\kappa_A=\kappa_B$, because the von Neumann entropy as the measure 
of entanglement is maximum when $\lambda_{1}= \lambda_{2} =1/2$. 
In addition, if the basis states are orthogonal, 
then it is easy to get complete entanglement.
Furthermore, S.van Enk showed [18] that {\it one  cannot generate entangled 
coherent state with complete entanglement from coherent states of independent 
two modes, and that one cannot generate entangled coherent state with 
complete entanglement from incomplete entangled coherent state by local 
operation involving local filtering}.

Let us return to the discussion of the attacks.  If Eve makes a scheme of 
the entanglement attack in front of the Bob's receiver, the states come from 
Alice are non orthogonal, because the signal power is very small. 
It becomes a problem of the generation of entangled state for non orthogonal states. 
Eve cannot get complete entanglement by her probe  according to the above statement. 
But she can send  a trick state which is generated by Eve herself. 
The amplitude of  Alice's coherent state is one of $M$, 
so Eve does not know the amplitude of signals. As a result, 
she has to prepare the entangled coherent states with $M$ amplitudes
The probability for getting the appropriate entangled coherent state  
is order of $1/M^2$, in addition she cannot know which is the complete one. 
Thus, the information on the quantum states of Eve involves inherent error.  
In addition, if the channel involves optical amplifier as repeater, 
then the received states are mixed state. 
Eve cannot make a good entanglement operation in this case.

If Eve makes the scheme at the Alice's side, 
then almost states are regarded as orthogonal, because the signal power is large.
Let us show a property of 
decoherence due to energy loss on the state 
$|\Psi_2\rangle_{AB}$ which has complete entanglement. 
Eve has the source of 
entangled state, keep one part and transmit the other part 
to Bob through a lossy channel. Bob will receive 
the attenuated optical state. This is a model for specific 
sharing method of entangled state.
In such a situation, Eve can prepare the following 
entangled state 
\begin{equation}
|\Psi^*_2(0)\rangle_{AB} 
=h^*_2( |\alpha\rangle_A |-\frac{\alpha}{\sqrt{\eta}}\rangle_B
-|-\alpha\rangle_A |\frac{\alpha}{\sqrt{\eta}}\rangle_B )
\end{equation} 
where $h^*_2=1/\sqrt{2(1-\kappa_A \kappa_B)}$, 
$\kappa_A = \langle \alpha | -\alpha \rangle$, and 
$\kappa_B = \langle \frac{\alpha}{\sqrt{\eta}} 
| -\frac{\alpha}{\sqrt{\eta}} \rangle$,
and where $\eta$ is transparency of the channel.
When we employ a half mirror model for energy loss channel, 
the effect of energy loss is described by a linear coupling 
with an external vacuum field as follows:
\begin{equation}
U_{BL}|\alpha\rangle_B|0\rangle_L = 
    |\sqrt{\eta}\alpha\rangle_B|\sqrt{1-\eta}\alpha\rangle_L. 
\end{equation}
where the mode $L$ is an external mode corresponding to 
energy loss, and $\alpha$ is taken as real. 
If we use $|\Psi^*_2(0)\rangle_{AB} $ as the initial state, 
Alice will finally couple with the environment in 
\begin{eqnarray}
&\hat I_A\otimes U_{BL}|\Psi^*_2(0)\rangle_{AB} 
\otimes |0\rangle_L& \nonumber\\
&=h_2(|\alpha\rangle_A 
       |-\alpha\rangle_B
       |-\sqrt{\frac{1-\eta}{\eta}}\alpha\rangle_L
      -|-\alpha\rangle_A 
       |\alpha\rangle_B
       |\sqrt{\frac{1-\eta}{\eta}}\alpha\rangle_L ).& 
\end{eqnarray}
Then the state shared by Eve and Bob is given by 
a super operator calculation  as follows:
\begin{eqnarray}
&\rho_{AB}= \tilde{h_2}^2
         \{\frac{1}{h^2_2}|\Psi_2\rangle_{AB} \langle\Psi_2|& 
         \nonumber\\
         &+(1-L)
         (|\alpha,-\alpha\rangle_{AB}\langle-\alpha,\alpha|
         +|-\alpha,\alpha\rangle_{AB}\langle\alpha,-\alpha|)\}&
\label{coherent_entangle}
\end{eqnarray}
where 
\begin{equation}
|\Psi_2\rangle_{AB}={1\over\sqrt{2(1-\kappa_A^2)}}
                ( |\alpha\rangle_A |-\alpha\rangle_B
                     -|-\alpha\rangle_A |\alpha\rangle_B ), 
\end{equation}
and where
$L = \exp\{-4(1-\eta)|\alpha|^2\}$, 
$|\alpha,-\alpha\rangle_{AB}=|\alpha\rangle_A |-\alpha\rangle_{B}$, 
and
$\tilde{h_2}^2=1/\{2(1-L\kappa_A^2)\}$.
As a result, we have the following form:
\begin{eqnarray}
&\rho_{AB}= \frac{(1-\kappa_A^2)}{(1-L\kappa_A^2)}
|\Psi_2\rangle_{AB} \langle\Psi_2|& \nonumber\\
         &+\frac{(1-L)}{(1-L\kappa_A^2)}
            (|\alpha,-\alpha\rangle_{AB}\langle-\alpha,\alpha|
            +|-\alpha,\alpha\rangle_{AB}\langle\alpha,-\alpha|)\}
\end{eqnarray}

Here let us discuss properties of entanglment of the above density 
operator. We can use the entangled 
fraction to measure the entanglement of mixed state.
The fully entangled fraction of this density operator is given by
\begin{equation}
f(\rho_{AB}) = _{AB} \langle \Psi_2 |\rho_{AB}|\Psi_2 \rangle_{AB}  
\end{equation}
because $|\Psi_2 \rangle_{AB}$ has complete entanglement 
which is independent of $\alpha$.
As a result, we have
\begin{equation}
f(\rho_{AB})=\frac{(1-\kappa_A^2)}{(1-L\kappa_A^2)}+
\frac{(1-L)(1-\kappa_A^2)^2}{(1-L\kappa_A^2)}
\end{equation}
In the special cases, the fraction of entanglement is that
$f(\rho_{AB})=1$ for $\eta=1$, and
$f(\rho_{AB})\ge \kappa_A^2(1-\kappa_A^2)$ for $\eta \cong 0$.
Although this is robust for energy loss compared 
with bi-photon entangled state, the entanglement 
is  destroyed when $\eta \cong 0$. So it is useless in the sense 
for getting any information.

\section{Quantum M-ary cipher for fiber network}
An implementation of Y-00 in the above was discussed based on phase 
(or polarization)modulation which provides the same energy for bit[11]. 
To apply them to fiber communication system, it is better to realize 
them by intensity modulation/direct detection or frequency shift keying. 
But to keep the security, it will require some additional idea for 
implementation of Y-00.
The essential role of modulation scheme in Y-00 is 
to assign non-commuting or same density operators to 0 and 1 for Eve.
In the original implementation[11], they assumed that the  selection 
of a set of the basis states to transmit 0 and 1 is as follows
\begin{equation}
\{1, 0 \}\rightarrow \{ |\Psi_1(1)\rangle, |\Psi_0(1)\rangle \}, 
\{ |\Psi_1(2)\rangle, |\Psi_0(2)\rangle \},
\dots
\end{equation}
where $\{ |\Psi_1(j)\rangle, |\Psi_0(j)\rangle \}, j\in \cal{M}$ 
is a set of basis states. The bit or key information (1 and 0 ) is transmitted 
by one of $M$ sets of basis states, controlling by initial shared key and running key.
This idea comes from 
a fundamental principle in quantum detection theory such that we can construct 
non-commuting density operators only by set of non-orthogonal states 
when one does not allow overlap of the selection of a set of basis state for 1 and 0. 
The non commutative nature of density operators depend 
on strongly on amplitude or energy of coherent state. 
In order to improve this effect, huge number of set of basis states must be prepared.

On the other hand, when one allows overlap for selection of a set of basis state, 
it is easy to construct the same density operators for 1 and 0. 
That is, $\rho_1=\rho_0$. 
{\it This scheme is called overlap selection keying}(OSK).
As an advantage of the overlap selection keying, we can say that it provides a very simple system 
and Eve completely cannot estimate the bits at that time 
even if the set of the basis states is only one. 
Each set of basis state is
used for $\{1, 0\}$, and $\{0, 1\}$, depending on running key.
\begin{eqnarray}
Set\quad A_1 : 0 &\rightarrow& |\alpha_{(1)}\rangle, 
\quad 1 \rightarrow |\alpha_{(M+1)}\rangle\\
Set\quad A_2 : 0 &\rightarrow& |\alpha_{(M+1)}\rangle, 
\quad 1 \rightarrow|\alpha_{(1)}\rangle
\end{eqnarray}
So the density operators of 1 and 0 for Eve are 
\begin{equation}
\rho_1=\rho_0 = \frac{1}{2}(|\alpha_{(1)}\rangle\langle \alpha_{(1)}|
+|\alpha_{(M+1)}\rangle\langle \alpha_{(M+1)}|)
\end{equation}
However, if the signal energy is large, the coherent states become of orthogonal. 
So bit strings that Eve observed are 
perfectly the same as those of Bob, though Eve cannot know information bit.
This gives still insecure situation.
In order to cope with the above problem,we employ
a combination of $M$-ary scheme and overlap selection keying. 
As a result, the energy dependency of the security may be alleviated, 
and it is enough to take into account only the  
average error probability for discrimination of 
2$M$ pure states for the security.

Here  we show an example of design for IMDD system. 
Let us assume that the maximum amplitude is fixed as $\alpha_{max}$.
We divide it into 2$M$. So we have $M$ sets of basis state
$\{(A_1,A_2),(B_1,B_2), \dots \}$. 
Total set of basis state is given
as shown in Fig.1. 
\begin{figure}
 \begin{center}
 \begin{tabular}{c}
 \includegraphics{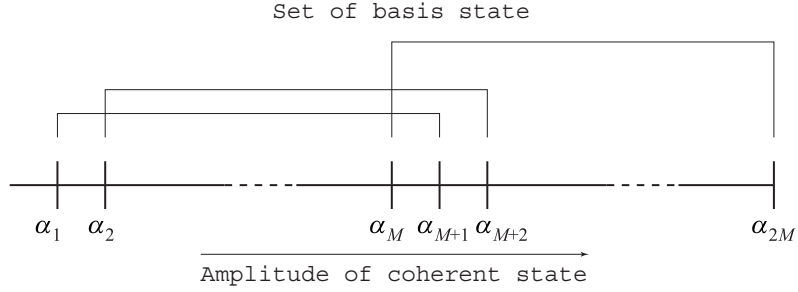}
 \end{tabular}
 \end{center}
\caption{Scheme of selection of quantum state parameters}
\label{figure1} 
 \end{figure}
Here, let us assign 0 and 1 by the same way as Eqs(19),(20) for the sets of $\{B_1,B_2\}$, 
$\{C_1,C_2\}$,$\dots $.
In this case, Eve completely cannot get bit information, while the knowledge of Eve 
depends on $\alpha_{max}$ and $M$ in the original scheme[11].
This fact gives a great advantage when we apply the Y-00 protocol 
to the practical system. 
Then she can try to know 
the information of quantum states used for bit transmission. 
This problem is of discrimination for 2$M$ pure states.
The error probability is given by Eq(4).
Although we have many results for calculation of quantum optimum detection 
problems, to give the analytical solution of this problem 
is still difficult at present time, 
because the set of states does not have complete symmetric structure. 
However, the tight upper bound is given by applying square root measurement(SRM) to 
2$M$ pure states.
It is well known that SRM provides the optimum or suboptimum detection 
for many kinds of pure state signals[19, 20]. The SRM is defined as follows:
\begin{equation}
|\mu_i \rangle=\hat{G}^{- \frac{1}{2}} |\psi_i \rangle , \quad \quad 
   \hat{G}= \sum\nolimits\limits_{\mit i\rm \in \cal{M}}^{} 
   |\psi_i \rangle \langle \psi_i |
\end{equation}
where $|\psi_i \rangle $ is signal state, 
$\hat{G}$ is the Gram operator, and the detection operator is formed as
\begin{equation}
\Pi_i=|\mu_i \rangle\langle \mu_i |, \quad i \in \cal{M}
\end{equation}
It is very easy to calculate the average error probability when 
we apply this detection operator, 
and these are very tight which can be proven by Helstrom's Bayes cost 
reduction algorithm[21] to check the tightness(see Fig.2). 
Thus if $M$ increases, then her error for information on quantum states increases. 
In this case, pure guessing corresponds to $\frac{2M-1}{2M}$. 
The lower bound is given by the minimum error probability:$P^*_e(2)$ 
for signal set $\{|\alpha_{(1)}\rangle,|\alpha_{(2)}\rangle \}$ 
which are neighboring states. It reads 
\begin{equation}
P_{e(2)}=\frac{1}{2}(1-\sqrt{1-\exp[-|\alpha_2-\alpha_1|^2]})
\end{equation}
For the error probability of Bob, if we can allow Bob 
to use the quantum optimum receiver, 
then the error probability of Bob is independent of the number of 
set of basis state, and it is given as follows:
\begin{equation}
P_e(B)= \frac{1}{2}(1 - \sqrt{1-|\langle \alpha_1|\alpha_{M+1}\rangle|^2})
\end{equation}
We emphasize that Eve cannot get bit information in this stage, 
because the information for 1 and 0 are modulated by the method of Eqs(19),(20).
We will show the detectability for the case of conventional 
receiver in the next section.
Thus, it is achieved 
that the error of Bob is less than that of Eve even the SNR of Eve 
is greater than that of Bob, which corresponds to the advantage distillation.
\begin{figure}
 \begin{center}
 \begin{tabular}{c}
 \includegraphics{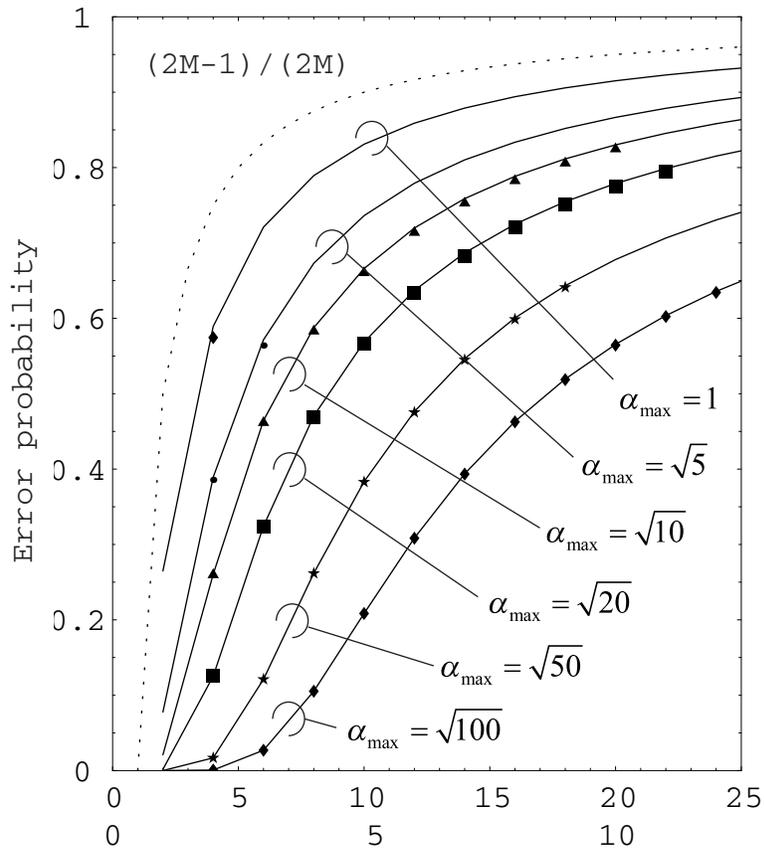}
 \end{tabular}
 \end{center}
\caption{
Error probability of Eve for discrimination of M pure states. 
Horizontal line: upper is of the number of state, under is the number of basis state}
\label{figure2} 
\end{figure}
\begin{figure}
 \begin{center}
 \begin{tabular}{c}
 \includegraphics{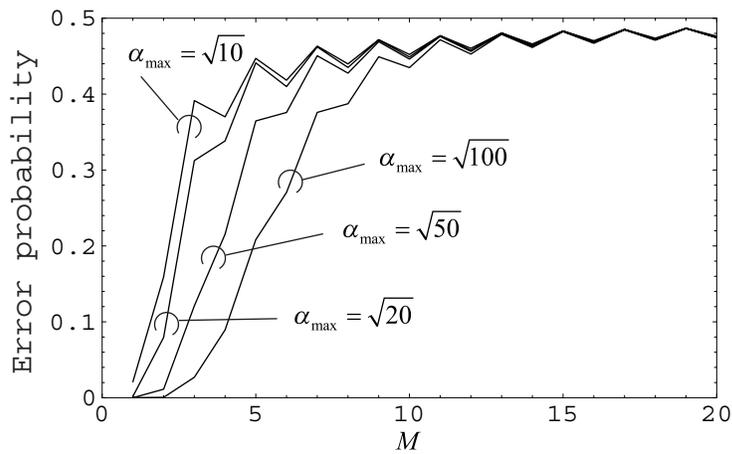}
 \end{tabular}
 \end{center}
\caption{Error probability of bit information for non overlap selection keying}
\label{figure3} 
\end{figure}
Let us compare with non-overlap and overlap selection keying.
In the former case, Eve will try to get bit information, 
so the density operators for Eve 
become mixed states $\rho_1$, and $\rho_0$ consisting of set 
of quantum states which send 1 and 0,
respectively. The error probability of Eve for bit information
($\rho_1$ or $\rho_0$) depends on
strongly power of laser light(see Fig 3).
In the latter case, the error of Eve is always 1/2, 
and it dose not depend on power of laser light.
On the information for set of basis states, the error probability of Eve
becomes the same one given by the formulae Eq(4) in both cases. 
Thus in the latter case, it is enough to take the information on 
the set of basis states into account.
Since, in the intensity modulation, we cannot keep the enough signal distance 
under the power constraint in comparison with phase modulation, the overlap selection keying
has great advantage.
In addition, it can send 2$M$ bits by $M$ sets of basis state. 
The above example was described as ASK(amplitude shift keying), but 
it is the same as the intensity modulation scheme.
As simple modification, we can use frequency shift keying(FSK) 
and also combination of ASK and FSK.
If one wishes more efficient scheme, then one can apply many modulation scheme, 
for instance Manchester scheme and so on.

\section{Design for practical fiber network}
A special feature of YK and Y-00 protocol is that one can design the secure 
system based on only 
the abilities of Eve and Bob for detectability. 
That is, it is not important 
how many repeaters are used in communication channel from Alice to Bob.
In general, the present fiber network consists of optical repeaters 
like fiber amplifier,
and direct detection scheme(intensity detection for on and off).
Although one can make the ability of Eve the level of pure guessing 
by increasing $M$(the number of basis set), it is difficult to improve 
the detectability of Bob when the communication distance is 
very long under the high bit rate, for example 1000 km and 1 Gbps.
Here we show how we can solve it.

A  practical fiber system consist of transmitter(laser diode), 
fiber cable, repeater(fiber amplifier), pre-amplifier, and photo-detector. 
The  noise in receiver of the above scheme of 
IMDD is given as follows:
\begin{equation}
\langle {I_{on}}^2 \rangle = \langle {I_{th}}^2 \rangle 
+\langle {I_{sig}}^2 \rangle +2\langle {I_{sp}}^2 \rangle 
+\langle {I_{sig-sp}}^2 \rangle +2\langle {I_{sp-sp}}^2 \rangle 
\end{equation}
where
$\langle {I_{th}}^2 \rangle,\langle {I_{sig}}^2 \rangle,\langle {I_{sp}}^2 \rangle ,
\langle {I_{sig-sp}}^2 \rangle, \langle {I_{sp-sp}}^2 \rangle $ are 
thermal noise in detector, quantum shot noise, noise by spontaneous 
emission from amplifier, beat noise between signal and spontaneous emission, 
beat noise between spontaneous emission itself, respectively. 
They are given as follows:
\begin{eqnarray}
\langle {I_{sig}}^2 \rangle&=&2e^2G_p\kappa_r\langle n \rangle B\\
\langle {I_{sp}}^2 \rangle &=&2e^2[\kappa_rG_pN(G-1)+(G_p-1)]n_{sp}B\delta f\\
\langle {I_{sig-sp}}^2 \rangle&=&4e^2G_p[\kappa_rG_pN(G-1)+(G_p-1)]
\kappa_r\langle n \rangle n_{sp}B\\
\langle {I_{sp-sp}}^2 \rangle&=&2e^2[\kappa_rG_pN(G-1)+(G_p-1)]^2
{n_{sp}}^2 B\delta f
\end{eqnarray}
where $e$ is the electron charge, $G_p$ is the gain of pre-amplifier,  
$G = \frac{1}{\kappa_r}$ is the gain of the repeater, $\kappa_r$ is the transparency
of the fiber, $N$ is the number of repeater,  $\langle n \rangle$ is 
the photon number per second at transmitter,
$n_{sp}$ is the spontaneous emission factor, $B$ is the bandwidth, 
and $\delta f$ is the bandwidth of optical filter, respectively.  
As a result, the bit sequence of Bob involves many error bits. 
That is, the error probability of Bob $P^*_e(B)$ in a real fiber system 
is not so good in comparison with quantum optimum receiver.
\begin{equation}
P^*_e(B) \gg \frac{1}{2}(1 - \sqrt{1-|\langle \alpha_1|\alpha_{M+1}\rangle|^2})
\end{equation}
Here we give 
a method for the improvement of this unavoidable error in the direct detection.
Since we can use high bit rate(Gbps), it does not give a degradation 
of the performance even if we use redundant codewords 
for sending bit information(0 and 1). In the following we show how it works.
Let us assign codeword:$u_i$ from the set $W:\{ u_i\in W\}$ 
of the sequence of quantum states. 
For example, as error correction code, we assign one codeword from the following 
set of codewords:
\begin{eqnarray}
& & Code_1:\{0\rightarrow u_1=|\alpha_{(1)}\rangle|\alpha_{(1)}\rangle
|\alpha_{(M+1)}\rangle, \quad
1\rightarrow u_4=|\alpha_{(M+1)}\rangle|\alpha_{(M+1)}\rangle|\alpha_{(1)}\rangle \}\nonumber\\
& &Code_2:\{0\rightarrow u_2=|\alpha_{(1)}\rangle|\alpha_{(M+1)}\rangle
|\alpha_{(1)}\rangle, \quad
1\rightarrow u_5=|\alpha_{(M+1)}\rangle|\alpha_{(1)}\rangle|\alpha_{(M+1)}\rangle \}\nonumber\\
& &Code_3:\{ 0\rightarrow u_3=|\alpha_{(M+1)}\rangle|\alpha_{(1)}\rangle
|\alpha_{(1)}\rangle \}, \quad
1\rightarrow u_6=|\alpha_{(1)}\rangle|\alpha_{(M+1)}\rangle|\alpha_{(M+1)}\rangle \}\nonumber\\
& &\vdots
\end{eqnarray}
If the set $A_2$ is chosen, then only bit 0 and 1 are changed, and 
the codeword is the same one.
The structure of this codeword may be open for public, 
but the selection depends on initial key and running key.

Let us recall that the decision of legitimate user is  binary.
For simplifying the evaluation, we assume that the probability distribution 
of noise in the detector for  bit is symmetric. 
That is, the conditional probabilities for 0 and 1 as the decision process 
are $P(1|0)=P(0|1)$.
Since the Hamming distance of this code is 3, it has a function of one bit 
error correction. So the average error probability is given as follows:
\begin{equation}
P_e = 3P(0|1)^2 - 2P(0|1)^3
\end{equation}
If the bit error:$P(0|1)=10^{-4}$ in the practical receiver of direct detection, 
then it turns the value of average error into about $10^{-8}$ 
which is enough for practical use.
This is effective rather than only to increase the transmitter 
energy of each light pulse for improving the error probability of Bob.

The security for bit information is kept in the case of overlap selection keying, 
because it is independent of signal power.
For information on the set of basis states which are used for sending bit sequence, 
Eve has to design 
the quantum optimum receiver applying to total number of pure quantum states 
as all kind of codewords  $u_i\in W$, because Eve cannot know them. 
In this case, the number of codeword is 3$M$.
This corresponds to increasing the set of basis state.
According to quantum detection theory and 
the numerical example of the error probability for $M$-ary pure states, 
the average error probability increases rapidly when the number $M$ increases.

In this section, we showed that if we insert key for the selection of 
codeword  into Y-00 protocol, 
then  we can improve the detectability of Bob without 
the degradation of the security.
More general discussion for design of error correcting code will be 
given in the subsequent paper.

\section{Conclusions}
We repeat our motivation here. Basically, one wants to realize 
one way secure communication. 
However, the situation of the eavesdropper is in general better 
than the legitimate user. To cope with such a situation, many proposals were 
devoted to discuss only a key distribution which allow two way 
communication or feedback to establish the advantage distillation. 
BB-84 and other quantum key expansion 
protocols provided a solution for such problems, introducing to 
detect the  existence of eavesdropper and privacy amplification. 

However,these may be unenterprising strategy in the practical system.
Following such well known results, Yuen verified that more simple 
information theoretical secure communication and unconditionally secure key 
expansion exist. 
In this paper, 
we have shown an efficient physical layer of Y-00 protocol applicable to  
secure communication and quantum key expansion for fiber system.
It was clarified that the overlap selection keying of set of basis states gives 
the very efficient way to establish the secure condition in Y-00 protocol.
This work is an extension of reference 22.

\section*{Acknowledgment}
We are grateful to H.P.Yuen, S.J.van Enk 
for helpful suggestions.


\end{document}